\documentclass[aps,prl,showpacs,twocolumn,groupedaddress]{revtex4}

\usepackage{graphics}
\usepackage{graphicx}
\usepackage{bm}

\begin{document}
\newcommand {\ve}[1]{\mbox{\boldmath \(#1\)}}
\newcommand {\ga}{\raisebox{-.5ex}{\(\stackrel{>}{\sim}\)}}
\newcommand {\la}{\raisebox{-.5ex}{\(\stackrel{<}{\sim}\)}}
\newcommand{\beq}{\begin{equation}}
\newcommand{\eeq}{\end{equation}}

\def\vec#1{{\bm #1}}

\title{\bf Velocity of vortices in inhomogeneous Bose-Einstein condensates}
\author{Halvor M. Nilsen,$^{a,b}$ Gordon Baym,$^{b,c}$ and C. J.
Pethick$^{b,c}$}
\affiliation{
$^{a}$Center for Mathematics for Applications, P.O. Box 1053 Blindern,
NO-0316   Oslo, Norway \\
$^{b}$NORDITA, Blegdamsvej 17, DK-2100 Copenhagen \O, Denmark\\
$^{c}$Department of Physics, University of Illinois at Urbana-Champaign,
   1110 West Green Street, Urbana, Illinois 61801-3080}

\date{\today}

\date{\today}

\begin{abstract}

    We derive, from the Gross-Pitaevskii equation, an exact expression for the
velocity of any vortex in a Bose-Einstein condensate, in equilibrium or not,
in terms of the condensate wave function at the center of the vortex.  In
general, the vortex velocity is a sum of the local superfluid velocity, plus a
correction related to the density gradient near the vortex.  A consequence is
that in rapidly rotating harmonically trapped Bose-Einstein condensates,
unlike in the usual situation in slowly rotating condensates and in
hydrodynamics, vortices do not move with the local fluid velocity.  We
indicate how Kelvin's conservation of circulation theorem is compatible with
the velocity of the vortex center being different from the local fluid
velocity.  Finally we derive an exact wave function for a single vortex near
the rotation axis in a weakly interacting system, from which we derive the
vortex precession rate.

\end{abstract}

\maketitle

\section{Introduction}

    Vortices in rapidly rotating trapped Bose-Einstein condensates
\cite{tk,tk1,Sinova,qhmodes} do {\it not} move with the local fluid flow,
despite Lord Kelvin's circulation theorem which implies that a vortex in a
perfect fluid generally cannot escape from a contour co-moving with the fluid
\cite{kelvin,Landau}.  To understand how vortices do in fact move in rapidly
rotating condensates, we derive here an exact expression for the velocity of a
vortex, applicable both to equilibrium and non-equilibrium situations
involving arbitrary numbers of vortices, through analyzing the motion of the
vortex singularity directly from the Gross-Pitaevskii (GP) equation
(Eq.~(\ref{GP}) below).  This differential equation, a generalization of the
Schr\"odinger equation, determines the time dependence of the order parameter
$\Psi$ (the condensate wave function) of the condensed system; its use in
studying vortex motion was pioneered by Fetter \cite{sandy-thesis}.  The great
advantage of the GP equation, as opposed to the hydrodynamic equations of a
perfect fluid, is that it provides a detailed model for the vortex core, and
may therefore be employed in the regime in which the vortex core size is
comparable to or greater than other length scales in the problem.

    We elucidate, via this approach, how the velocity of the center of a
vortex and the local fluid velocity differ when the density of the condensate
varies sufficiently rapidly in the neighborhood of a vortex -- the situation
in rapidly rotating condensates.  Furthermore, we show how the difference
between the vortex velocity and the fluid flow in such cases is compatible
with Kelvin's theorem, a result with more general applications in
hydrodynamics.

\section{The velocity of a vortex}

    The difference between the local fluid velocity and the velocity of the
center of the vortex is brought out very clearly by the simple example of a
single off-center vortex at radius $b$ in a two-dimensional harmonic trap,
$V({\vec r})$, of frequency $\omega$, in the limit in which the interparticle
interaction is negligible.  Such a system is described by a wave function,
$\Psi$, that is a linear superposition of the (s-wave) oscillator ground state
and a p-wave eigenstate of the trapping potential.  At time $t=0$,
$\Psi(\zeta,\zeta^*) \sim (\zeta-b) e^{-|\zeta|^2/2d^2}$, in the usual complex
notation in which $\zeta = x+iy$; here $d=\sqrt{\hbar/m\omega}$ is the
oscillator length, and $m$ the particle mass.  Since the oscillator ground
state has energy $\hbar\omega$ and the p-state has energy $2\hbar\omega$, the
time dependent wave function is
\beq
  \Psi(\zeta,\zeta^*,t) \sim
  (e^{-2i\omega t}\zeta  -e^{-i\omega t} b ) e^{-|\zeta|^2/2d^2}.
\label{wave}
\eeq
At time $t$ the vortex is located at $\zeta=be^{i\omega t}$, and thus its
center precesses in the positive sense about the origin at the trap frequency.
On the other hand, the fluid velocity, given by $(\hbar/m) \nabla \phi$, where
$\phi$ is the phase of $\Psi$, equals $\hbar /m\rho$ around the vortex, where
$\rho$ is the distance from the vortex.  There is no background fluid flow;
the only flow is that due to the vortex itself, and yet the vortex precesses
with frequency $\omega$.  A naive application of Kelvin's theorem would
suggest that the vortex should be stationary.

    We turn now to a more general analysis of the motion of singly quantized
vortices in two dimensions, as described by the time-dependent GP equation,
\beq
  i\hbar \partial \Psi/\partial t
 = -(\hbar^2/2m)\nabla^2\Psi + V\Psi +g_2|{\Psi}|^{2} \Psi,
 \label{GP}
\eeq
where $V(\vec r\,)$ is the transverse trapping potential, and $g_2$ is the
two dimensional coupling constant.  We normalize $\Psi$ by $\int d^2 r|\Psi^2|
= N$, where $N$ is the total number of particles.  For a system that is
uniform in the axial direction, $g_2 = 4\pi\hbar^2 a_s/mZ$, where $a_s$ is the
s-wave interatomic scattering length, and $Z$ is the height of the system.
Given $\Psi$, the left side, evaluated at the position of the vortex tells us
the velocity of the vortex position.  Thus the instantaneous velocity of the
vortex center depends only on the value of $\nabla^2\Psi$ at the vortex
position.  The simplest case is that of a singly quantized cylindrically
symmetric vortex at position $\zeta_i$ in a spatially uniform system, which is
described by a wave function $\Psi(\zeta)$, $\sim\zeta-\zeta_i$ close to the
vortex.  More generally, the wave function of an asymmetric vortex, e.g., in
an elliptic container, can include a term $\sim (\zeta-\zeta_i)^*$ as well.
Thus, to evaluate the GP equation at the vortex position, we can write the
wave function in the neighborhood of a singly quantized vortex at position
$\zeta_i$, without loss of generality, as
\begin{equation}
 \Psi(\zeta,\zeta^*,t)=\left\{\left[\zeta-\zeta_i(t)\right] +
  \alpha\left[\zeta-\zeta_i(t)\right]^*\right\}e^{-Q+i\phi_b},
 \label{Wavefcn}
\end{equation}
where $|\alpha|<1$ for a vortex with positive circulation, and
$Q(\zeta,\zeta^*,t)$ and $\phi_b(\zeta,\zeta^*,t)$ are real, smooth functions
at $\zeta=\zeta_i(t)$.  The background phase, $\phi_b$, is that of the
superfluid with the singular contribution from the vortex at $\zeta_i$
removed, while the function $Q$ describes the background density variation
at the vortex.  The background fluid velocity is $v_b \equiv
v_{bx}+iv_{by} = 2(\hbar/m)\partial \phi_b/\partial \zeta^*$.  Then the GP
equation implies
\begin{equation}
  \frac{\partial}{\partial t}\zeta_i  +\alpha \frac{\partial}{\partial
     t}\zeta_i^*
   = \frac{2\hbar}{m}\left(\frac{\partial}{\partial \zeta^*}+\alpha
    \frac{\partial}{\partial \zeta}\right)(\phi_b+iQ)\big|_{\zeta_i}.
  \label{vv_complex}
\end{equation}
This equation shows how the instantaneous velocity of the vortex center
depends only on the slope and curvature of the wave function in the
neighborhood of the vortex core; the velocity depends on the trapping
potential, $V$, and the interaction strength, $g_2$, only insofar as they
affect the value of the gradient of $Q$ at the vortex position.

    After straightforward algebra we find that the velocity, ${\vec u}_i$, of
the center of the vortex at $\zeta_i$ is, in Cartesian coordinates,
\begin{eqnarray}
  u_{ix} &=&  v_{bx}
     -\frac{\hbar}{m{\rm Re}(\lambda^{-1})}\left( \frac{\partial
   Q}{\partial y}
          + {\rm Im}(\lambda^{-1})\frac{\partial Q}{\partial x}\right),
           \nonumber \\
  u_{iy}  &=& v_{by}
     +\frac{\hbar}{m{\rm{Re}}\lambda}\left(\frac{\partial Q}{\partial x}
          - {\rm Im}\lambda\frac{\partial Q}{\partial y}\right),
 \label{vel}
\end{eqnarray}
where $\lambda = (1-\alpha)/(1+\alpha)$, and all quantities on the right
are evaluated at $x_i,y_i$.  For $\alpha = 0$,
\beq
    {\vec u}_i = \left({\vec v}_b +\frac{\hbar}{m}\hat{\vec z}\times \nabla
        Q\right)_{x_i,y_i}.
 \label{alpha0}
\eeq
The vortex velocity is the sum of the local fluid velocity, ${\vec v}_b$,
plus a correction, $\sim \nabla Q$, of order $(\hbar/m)\nabla n_b $, where
$n_b \sim e^{-2Q}$ describes the background density in the vortex core region.
This correction is non-negligible if the background density varies
significantly over the vortex core.  In the simple example above, $v_b=0$ and
the entire vortex velocity arises from the density gradient.  As this
calculation illustrates, vortices do not in general move with the local fluid
velocity, generated, e.g., by other vortices in the system.  For the simple
example in Eq.~(\ref{wave}), $\phi_b=0$, while $\nabla Q = \vec r/d^2$, and
$\vec u_i = \omega(\hat z \times \vec r_i)$.  We note that these results hold
for generalizations of the GP equation in which the interaction energy per
particle is given by a local function of the density, as in hydrodynamics.

\section{How Kelvin's theorem is satisfied}

    How can Eq.~(\ref{vel}) be reconciled with Kelvin's theorem, that the
circulation, $\oint_{\cal C} {\vec v}\cdot d{\vec s}$ around a contour, ${\cal
C}$, moving with the local fluid velocity, ${\vec v({\bf r})}$, is conserved
in time?  (Here $d\vec s$ is the line element.)  Kelvin's theorem applies to
the GP equation, as well as the Schr\"odinger equation, since the flow
described by these equations is potential.  (The proof of Kelvin's theorem for
the GP equation is given by Damski and Sacha \cite{damski}.  As stressed in
Ref.~\cite{damski} and earlier in Ref.~\cite{koplik} the conditions for the
proof of the theorem do not apply should the velocity be singular on the
contour.  This situation does not arise in the cases we consider here.)  Since
the theorem implies that a vortex cannot escape from a co-moving contour, no
matter how small, the vortex would appear to be constrained to move with the
background local fluid velocity.  However, as is illustrated in Fig. 1 for an
off-center vortex in the harmonic oscillator potential, co-moving contours do
not necessarily remain regular in time, but rather, due to the difference
between the fluid motion about the vortex and the motion of the vortex center
can become highly distorted.  While in this example, the flow velocity
calculated from (\ref{wave}) is always circular about the vortex center, the
density gradient term in the vortex velocity (\ref{alpha0}), which makes the
vortex move with respect to the background flow, causes a mismatch between the
motion of the contour and the motion $|v|=b\omega$ of the vortex center.  An
initially circular contour around the vortex begins to develop a lobe towards
larger radii from the center.  With time the lobe becomes extended and begins
to wrap around.  In Fig.~1 we show the fate, calculated numerically, of an
initially circular contour centered on the vortex, at successive times within
a quarter period, $\pi/2\omega$.  The standard argument that the vortex moves
with the local fluid velocity assumes that the average velocity on the contour
equals the fluid velocity at the center of the vortex.  As this calculation
indicates, it is not possible in general to deduce the motion of the vortex
from Kelvin's theorem when the average velocity on the contour is no longer
simply related to the vortex motion.

\begin{figure}[htb]
{\centering \includegraphics[width=8cm]{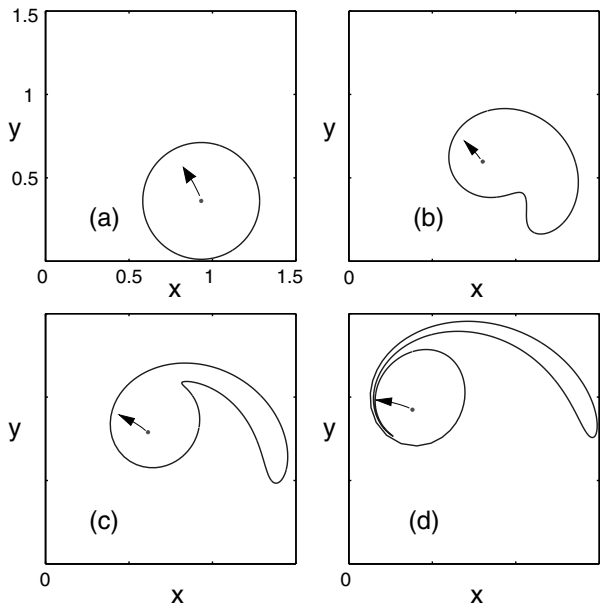}
\caption{
Contour co-moving with the fluid, surrounding a
single off-center vortex in a harmonic trap.  The contour is shown in the
stationary lab frame at successive equally spaced times, (a)-(d), within
one-quarter of an oscillation period, $\pi/2\omega$.  Distances are measured
in units of the oscillator length, $d$.  The vortex is at radius $d$ and the
initial radius of the contour is $0.3d$.  Note the sharp fold of the contour
in pane (d).}}
\label{contour_evolution}
\end{figure}

\section{Applications}

    We now apply the result Eq.~(\ref{vel}) to discuss more general vortex
motion in detail in two limiting cases.  The first is when the density scale
height $R$ is large compared with the vortex core size, determined in this
case by the healing length, $\xi$; such a situation arises, e.g., in a system
slowly rotating at angular frequency $\Omega$, when the interaction energy,
$\sim g_2n$ (where $n$ is the mean two-dimensional density), dominates the
rotational energy, $\hbar\Omega$, a strongly interacting regime in this sense.
The second is the weakly-interacting rapidly-rotating lowest Landau level
(LLL) limit in a harmonic trapping potential, where $g_2n\ll \hbar\Omega$.  In
the former case a vortex does move with the local fluid velocity, but as the
harmonic oscillator example indicates, it does not in the latter case.  The
motion of vortices in the limit of large $g_2n$ may be explicitly calculated
by the method of asymptotic matching \cite{asymp,pismen,svid,svid1,anglin},
which we recall here.  In this approach one solves the GP equation near the
vortex by expanding the wave function about the solution for a uniform system
to first order in the gradient of the potential.  In the far field, at
distances large compared with $\xi$, the velocity is determined by solving the
continuity equation with the density taken to be that in the absence of the
vortex.  Matching of the inner and outer solutions in the region where the
radial coordinate $\rho$, measured with respect to the position of the vortex,
is in the range $\xi \ll \rho \ll R$, gives the vortex velocity and the
parameters of the core wave function.  The solution near the vortex core,
which enters Eq.~(\ref{vel}), has the form \cite{asymp,pismen,svid,svid1}
\beq
\Psi=\left(|\Psi_0(\rho)| +\chi(\rho)\cos\theta\right)
  e^{i\theta+i\eta(\rho)\sin\theta}, \label{ansatz}
\eeq
where $\Psi_0$ is the vortex solution in a uniform medium, $\eta$ and
$\chi$ are real functions of $\rho$, and the azimuthal angle $\theta$, about
the center of the vortex, is zero in the outward direction (opposite to the
background density gradient near the vortex).  Comparing this structure with
the general form (\ref{Wavefcn}), we see that $\alpha$ vanishes in this limit
and the corrections due to asymmetry of the vortex core are negligible; the
background phase is $\phi_b= \eta(\rho)\sin\theta$.  Using the analysis of
Refs.~\cite{asymp,pismen,svid,svid1}, we find $\chi \sim \rho^4$ at small
$\rho$, and $\eta \sim \rho$.  In addition, for small $\rho$, $Q = -\chi\cos
\theta/|\Psi_0| \sim \rho^3\cos\theta$.  Thus, as we see from
Eq.~(\ref{alpha0}), the only contribution to the vortex velocity comes from
the background phase of the wave function; the vortex moves with the
background flow velocity.  The background fluid velocity, found from the
far-field solution \cite{svid,svid1,Lundh,Sheehy04}, is given to logarithmic
accuracy by \beq {\vec v}_b(\rho) = \frac{\hbar}{2m}\left(\hat{\vec z}\times
\nabla \ln n \right)\ln\left(\frac{\rho}{R}\right), \label{pr} \eeq in the
region $\rho \gg \xi$, while in the core region, $\rho/R$ should be replaced
by $\xi/R$.  Here $n$ is the smooth (Thomas-Fermi) density outside the vortex
core, given by $\nabla n = - \nabla V(r)/g_2$.  The flow velocity is induced
by the vortex as a consequence of the density gradient in the system.

    In the rapidly rotating limit, on the other hand, in which the size of the
vortex core is comparable to other lengths in the problem, vortices do not
move with the background flow.  In a harmonic trap the wave function in this
limit is, to a first approximation, a linear superposition of lowest
Landau levels (LLL) in the Coriolis force \cite{Jason01,Gentaro04}, $\Psi \sim
\prod_j (\zeta-\zeta_j) e^{-|\zeta|^2/2d^2}$.  Thus $\alpha=0$, and the
velocity of the individual vortices, Eq.~(\ref{vv_complex}), is given by,
\begin{equation}
  \frac{\partial}{\partial t}\zeta_i
   = i\frac{\partial}{\partial \zeta^*}(\omega|\zeta|^2)= i\omega \zeta_i,
  \label{LLL}
\end{equation}
i.e., $\vec{u}_i = \omega {\hat{z}}\times \vec{r}_i$.  In other words,
each vortex precesses at the trap frequency, independent of the flow from
other vortices!  This result implies immediately that in any motion of the
vortices other than precession at the trap frequency, as for example in
Tkachenko modes in the rapidly rotating limit \cite{tk}, higher Landau levels
play a crucial role in the wave function.

    As a first step in describing the effects of interactions on vortex motion
we consider the problem of steady state precession, in which there exists a
frame rotating at angular velocity $\Omega$ in which the system is stationary.
The problem then is to determine $\Omega$ as a function of the interaction
strength.  In general, $\Omega = \partial\langle H\rangle/\partial \langle L
\rangle$, where $L$ is the $z$ component of the angular momentum.  In a system
composed primarily of lowest Landau levels, $\langle H\rangle = \omega (
\langle L \rangle +N) + \langle H_{\rm int} \rangle$, so that $\Omega - \omega
= \partial\langle H_{\rm int}\rangle/\partial \langle L \rangle$.  As the
angular momentum of the system increases, the particles spread out, lowering
the average density; thus the derivative of the interaction energy with
respect to the angular momentum is negative, implying that $\Omega < \omega$.
The first deviation of $\Omega$ from $\omega$ is of order $g_2$.

    Instead of calculating $\Omega$ from $\partial\langle H\rangle/\partial
\langle L \rangle$, we discuss the precession of a single vortex (an effect
measured, in the small core regime, by Anderson et al.  \cite{JILAexpt}),
directly in terms of Eq.~(\ref{vel}).  Explicitly, we calculate the exact wave
function of a single vortex to order $b$ and $g_2$, and from this show how the
known vortex precession rate emerges for a vortex at small distance, $b$, from
the origin, to first order in the coupling strength \cite{linn}.  (We set
$\hbar$=1 in this section).  In the frame rotating at $\Omega$, the GP
equation assumes the form,
\beq
   (H_0-\Omega {\hat l} - \mu)\Psi = -g_2|\Psi|^2\Psi,
 \label{GPOmega}
\eeq
where $H_0$ is the two-dimensional oscillator Hamiltonian, ${\hat l}=
(\zeta\partial/\partial \zeta - \zeta^*\partial/\partial \zeta^*)$ is the
angular momentum operator, and $\mu$ is the chemical potential in the rotating
frame (equal to $\omega$ in the absence of interactions).  Since
$\Omega-\omega$ is first order in $g_2$, as is $\mu-\omega$, it follows from
Eq.~(\ref{GPOmega}) that the interaction term mixes LLL components of order
$g_2^0$ in $\Psi$ for all angular momentum $\nu$.  In the sense of
quantum-mechanical perturbation theory, while the interaction is of order
$g_2$, the splitting of the LLL and thus energy denominators are also of order
$g_2$, leading to $g_2^0$ corrections.

    To calculate the vortex velocity from Eq.~(\ref{vv_complex}), or
equivalently, (\ref{vel}), one must also include the higher Landau level
contributions to $\Psi$ explicitly; in general the exact wave function
describing a vortex at position $\zeta_0$ in terms of Landau levels is:
\beq
   \Psi(\zeta) =
  \sum_{\nu\sigma}C_{\nu\sigma}\left[P_{\nu\sigma}(\zeta,\zeta^*)
  -P_{\nu\sigma}(\zeta_0,\zeta_0^*)\right]e^{-|\zeta|^2/2d^2},
 \label{Psizeta}
\eeq
where the normalized Landau level wave functions are
$\chi_{\nu\sigma}(\zeta) = P_{\nu\sigma}(\zeta,\zeta^*)e^{-|\zeta|^2/2d^2}$,
with $P_{\nu\sigma}$ a polynomial, $\nu=0,\pm1,\pm2,\dots$ is the angular
momentum, and $\sigma=0,1,2,\dots$ is the radial quantum number.  The state
$\chi_{\nu\sigma}(\zeta)$ is an eigenstate of $H_0$ with energy
$(2\sigma+|\nu|+1)\omega$.

    The GP equation for a wave function stationary in the rotating frame,
evaluated at the vortex position $\zeta=b$, is
\beq
   \Omega \hat l  \Psi= -\frac{2}{m}\frac{\partial^2}{\partial
   \zeta\partial\zeta^*}\Psi.
\eeq
With $\Psi$ given by Eq.~(\ref{Psizeta}), this equation reduces to
\begin{eqnarray}
 (\Omega-\omega)\sum_{\nu\sigma}C_{\nu\sigma}
 \zeta\frac{\partial}{\partial \zeta} P_{\nu\sigma} \hspace{100pt}
 \nonumber\\
 = \sum_{\nu\sigma}C_{\nu\sigma}\left[(\Omega+\omega)\zeta^*
  \frac{\partial}{\partial \zeta^*}
  -\frac{2}{m}\frac{\partial}{\partial\zeta\partial\zeta^*}\right]
  P_{\nu\sigma},
\label{magic}
\end{eqnarray}
with all terms evaluated at $b$.  The $\omega$ terms arise from $\partial
e^{-|\zeta|^2/2d^2}/\partial\zeta\partial\zeta^*$.  Since $\Omega-\omega$ is
of order $g_2$ we need keep only LLL terms, $\sigma=1$ in the sum on the left,
and of these only the $\nu=1$ term is of order $b^0$.  On the right side, only
higher Landau levels enter, since the $P_{\nu0}$ are independent of
$\zeta^*$.

    The degree of mixing of higher $\nu$ in the wave function of a single
vortex depends on its distance $b$ from the origin.  To leading order in $b$,
$C_{\nu0} \sim b^{\nu-1}$ for $\nu\ge 1$, while $C_{00} \sim b$.  This
relation is readily proven recursively:  the interaction mixes terms
$\chi_{\nu 0}$, $\chi_{10}$, and $\chi_{00}^*$ to give a $\nu+1$ component;
since $C_{00}$ begins at order $b$, then if $C_{\nu0}$ is of order
$b^{\nu-1}$, $C_{\nu+1,0}$ begins at order $b^\nu$; other mixings that give a
$\nu+1$ component lead to terms at least of order $b^\nu$.  An immediate
consequence of this mixing is that even if one starts in the absence of
interactions with a single vortex at $b$, the wave function in the presence of
interactions describes further vortices at large distances, $\sim d^2/b$; in
general, the LLL part of $\Psi$ is proportional to a polynomial in $\zeta$ of
the form $\zeta f(b\zeta/d^2) - b$, to leading order in $b$, where $f(0)=1$,
so that in addition to the zero at $\zeta=b$, the polynomial has further
zeroes at $|\zeta|\sim d^2/b$.  However, the particle density is negligible at
such distances.

    As an explicit illustraion, let us start with a single vortex at position
$b$ described in the absence of interactions by
\beq
  \Psi(\zeta) = \sqrt{\frac{N}{\pi}}\frac{\zeta-b}{d^2}e^{-|\zeta|^2/2d^2},
\eeq
then with interactions the LLL wave function -- even to order $g_2^0$ --
includes levels $(\nu,\sigma=0)$ with amplitude $\sim b^{\nu-1}$ for $\nu \ge
1$.  To first order in $b$, we must include as well the $\nu=2,\sigma=0$ LLL.
Furthermore, the interactions to first order in $g_2$ mix in higher Landau
levels with $\nu=0,1,2$.  However the $\nu=2$ term contributes to the equation
for the motion of the vortex only in order $b^2$, and thus can be neglected.
Computing the coefficients $C_{\nu\sigma}$ by expansion of the GP equation
(\ref{GPOmega}) one readily finds the exact wave function of a single vortex
at position $b$, to first order in $b$ and $g_2$:
\begin{eqnarray}
 \Psi(\zeta) =
  \frac{C}{d^2}(\zeta-b)\left\{1 +\frac{b\zeta}{2d^2}
  +\frac{mg_2N}{16\pi \hbar^2} (|\zeta/d|^2-2) \right.
  \nonumber\\
 \left.
  + \cdots \right\}e^{-|\zeta|^2/2d^2},
 \label{wf}
\end{eqnarray}
where $C\sim\sqrt{N/\pi}$ is a normalization constant.  The term $\sim
\zeta(|\zeta/d|^2-2)$ is the $\nu=1,\sigma=1$ contribution.  (We include the
higher order $b^2$ term only to indicate that $\Psi(\zeta)$ has a zero at
$\zeta=b$.)

    Noting that $\alpha=0$ here, we substitute this wave function into
Eq.~(\ref{vv_complex}) (or equivalently, into Eq.~(\ref{alpha0})) to
derive the precession rate to first order in $g_2$ and lowest order in $b$,
\cite{linn}:
\beq
  \Omega-\omega = - \frac {Ng_2}{8\pi \hbar d^2}.
\eeq
This result illustrates the role of higher Landau levels in producing
deviations of the precession frequency from the trap frequency.  While the
background velocity, $\vec v_b$, vanishes in the non-interacting case, we see
in the presence of interactions, from Eq.~(\ref{alpha0}), that $\vec v_b=
b\omega \hat y/2$, for a vortex centered at $(x=b, y=0)$, while the second
term in (\ref{alpha0}) equals $(1-mgN/4\pi\hbar^2)b\omega \hat y/2$.  The
precessional motion arises in part from the background flow and in part from
the density gradient term.

    In addition, the chemical potential shift is given by $\mu-\hbar\omega =
3g_2N/8\pi d^2$, to first order in $g_2$ and lowest order in $b$.  To derive
the $b^2$ corrections to the precession rate and chemical potential would
require including the $\nu = 3$ components in $\Psi$.

\section{Summary}

    In summary, we have derived a general expression for the velocity of the
center of a vortex in an atomic Bose-Einstein condensate in terms of the
background fluid velocity and the density gradients at the vortex center.  The
differences of the two velocities becomes important when the vortex core size
becomes comparable to the scale height of the density variation.  In
particular, in the case of a rapidly rotating condensate in a harmonic trap,
vortices precess at the transverse trap frequency, $\omega$, minus corrections
of order $Na_s/Z$.

    We are grateful to David Thouless for insightful comments that led us to
use the GP equation directly to study the motion of vortices, to Benny Lautrup
for valuable discussion, and Sandy Fetter and Bogdan Damski for useful
correspondence.  This research was supported in part by NSF Grants PHY03-55014
and PHY05-00914.

\end{document}